\documentclass[reprint,amsmath,amssymb,aps,superscriptaddress]{revtex4-1}

\usepackage{bm}
\usepackage{color}
\usepackage{comment}

\usepackage{braket}

\usepackage{graphicx}
\usepackage{dcolumn}
\usepackage{here}

\usepackage{hyperref}
\hypersetup{
     colorlinks   = true,
     linkcolor    = blue,
     citecolor    = blue,
     urlcolor     = blue
}

\usepackage{appendix}

\newcommand{\R}{\mathrm{Re}}
\newcommand{\I}{\mathrm{Im}}
\newcommand{\A}{\mathcal{A}}

\newcommand{\x}{\bm{x}}
\newcommand{\kvec}{\bm{k}}

\newcommand{\q}{\bm{q}}
\newcommand{\J}{\bm{J}}

\begin{document}

\preprint{APS/123-QED}

\title{Orbital Gravito-Magnetoelectric response and Orbital magnetic quadrupole moment correction}

\author{Koki Shinada}
\email{shinada.koki.64w@st.kyoto-u.ac.jp}
\author{Robert Peters}
\affiliation{Department of Physics, Kyoto University, Kyoto 606-8502, Japan}

\date{\today}

\begin{abstract}
The magnetoelectric effect has been actively studied in multiferroics since the first observation in an antiferromagnetic, $\mathrm{Cr_2O_3}$. 
This effect appears in systems without spatial inversion symmetry and time-reversal symmetry and is sensitive to detecting magnetic quadrupole moments. 
It is often discussed as inducing spin magnetizations; however, the orbital magnetoelectric effect in metals has recently attracted much attention since its observation in $\mathrm{MoS_2}$ and twisted bilayer graphene. In this work, we propose the full quantum formalism for the temperature gradient induced-orbital magnetoelectric effect (orbital gravito-ME effect). The effect consists of two parts, i.e., an extrinsic part and an intrinsic part. We demonstrate that the intrinsic part needs a correction from the orbital magnetic quadrupole moment besides the usual Kubo formula to avoid an unphysical divergence at zero temperature and to satisfy the Mott relation.
Furthermore, we show the classification table with the magnetic point group for the intrinsic and extrinsic effects. Finally, we analyze the intrinsic part in a $\mathcal{PT}$-symmetric model exhibiting an orbital magnetization order, i.e., a loop current order, and demonstrate the enhancement near Dirac points. We believe that these results will contribute to the detection and usage of orbital magnetic moments beyond spin moments.
\end{abstract}

\maketitle

\section{Introduction}
Orbital angular momentum is one of the fundamental degrees of freedom of electrons along with charge and spin. 
Electrons bound to a nucleus form electron clouds, characterized by the orbital angular momentum such as $s$-wave symmetry and $p$-wave symmetry.
In solids, electrons are exposed to the periodic potential created by the lattice of nuclei, where some electrons are not bound anymore to a particular nucleus but can move through the lattice.
These electrons can circulate in solids generating orbital magnetic moments (OMMs) and can even create a bulk magnetization, such as in loop current order. 
The calculation of the OMM in solids has been a challenging theoretical problem because the angular momentum depends on the position operator that is unbounded and thus ill-defined in solids. This problem has been solved in the modern theory of the orbital magnetic moment \cite{PhysRevLett.95.137205,PhysRevB.74.024408,PhysRevLett.95.137204,PhysRevLett.99.197202,vanderbilt2018berry}.

The OMM is the cause of various phenomena. For example, the valley Hall effect has been proposed and experimentally observed as an analog to the spin Hall effect \cite{PhysRevLett.99.236809,doi:10.1126/science.1250140}. Here, orbital magnetic moments with opposite signs flow in opposite directions and accumulate at the edges. 
These phenomena imply the possibility of applying the OMM in magnetic devices, which is actively studied in valleytronics and orbitronics beyond spintronics. One goal here is to build a magnetic random access memory (MRAM) for writing data, where the magnetization can be controlled electronically.
Two mechanisms have been widely discussed for building MRAM; spin transfer torque (STT) and spin-orbit torque (SOT), where injecting a spin current switches the magnetization. Recently, a novel mechanism using the orbital current (orbital transfer torque) has been experimentally observed in $\mathrm{WTe_2/Fe_3GeTe_2}$ heterostructures \cite{Ye_2022}.

Besides the necessity of electrically manipulating magnetic moments to apply them in devices, there also is an ongoing fundamental interest in it. 
The magnetoelectric effect (ME effect) has been energetically studied in systems without inversion and time-reversal symmetry.
In the ME effect, magnetization can be induced by an electric field, or polarization can be induced by a magnetic field, and such cross-correlations are attracting attention for their ability to characterize the symmetry of materials. 
Since its first observation in the antiferromagnet $\mathrm{Cr_2O_3}$ \cite{1571980074033746432,astrov1960magnetoelectric,PhysRevLett.6.607}, the ME effect has been discussed mainly for spin degrees of freedom in multiferroics \cite{fiebig2005revival,wang2010multiferroic,tokura2014multiferroics,dong2015multiferroic,fiebig2016evolution}.
Magnetization can also be induced by a current, which  is called the Edelstein effect \cite{EDELSTEIN1990233}.
Research on the Edelstein effect has also focused on inducing a spin polarization in systems with strong spin-orbit coupling.
Of course, besides spin polarization, OMMs also generate magnetization. 
Thus, the orbital version of the ME effect and the Edelstein effect can be anticipated.
In fact, the orbital ME effect has been discussed in topological insulators. This effect has two parts: Kubo terms and a  Chern-Simons term \cite{PhysRevB.78.195424,PhysRevLett.102.146805,Malashevich_2010,PhysRevB.82.245118,PhysRevLett.112.166601}.
Especially, the Chern-Simons term has a topological nature and yields a quantized value for the ME effect in topological insulators. Furthermore, this term introduces the axion coupling rewriting the Maxwell equation anew and paving the way for the axion electrodynamics \cite{doi:10.1063/5.0038804}. However, the Chern-Simons term is not yet directly observed but only indirectly \cite{doi:10.1126/science.aaf5541,okada2016terahertz,liu2020robust,gao2021layer}.

While the orbital ME effect in topological insulators is diligently researched due to its topological property, the orbital ME effect in metals is just beginning to attract attention. The orbital Edelstein effect in metals has been theoretically discussed \cite{yoda2015current,PhysRevLett.116.077201,PhysRevB.102.184404,PhysRevB.102.201403,PhysRevB.96.035120} and recently observed in strained $\mathrm{MoS_2}$ \cite{lee2017valley,PhysRevLett.123.036806} and twisted bilayer graphene \cite{doi:10.1126/science.aaw3780,he2020giant}. 
In particular, twisted bilayer graphene has a large OMM, approximately proportional to the Berry curvature, reaching a magnitude of tens of Bohr magnetons. 
The orbital Edelstein effect is also discussed in superconductors yielding a non-dissipative response \cite{PhysRevResearch.3.L032012,PhysRevLett.128.217703}. 
Furthermore, the intrinsic orbital ME effect in metals is recently formalized using semiclassical theory \cite{PhysRevB.103.045401,PhysRevB.103.115432} or a fully quantum mechanical analysis \cite{shinada2022}.
It has been shown that the orbital ME effect can be used to detect $\mathcal{PT}$-symmetric orbital magnets, such as an antiferromagnetic loop current order, and it implies that the orbital ME effect is useful for the detection of higher-order multipoles such as orbital magnetic quadrupole moments. 

Besides electric fields, also temperature gradients can induce magnetization. The thermal version of the spin ME effect (spin gravito-ME effect) and the spin Edelstein effect are discussed in systems with spin-orbit coupling \cite{WANG20101509,PhysRevB.87.245309,xiao2015thermoelectric,PhysRevB.98.075307,PhysRevB.99.024404}. However, in theoretical calculations of thermal response functions, there are certain aspects that need special consideration. In general, when calculating the intrinsic thermal response, the Kubo formula alone is not sufficient; certain equilibrium contributions of magnetic multipoles need to be subtracted \cite{LSmrcka_1977,PhysRevLett.107.236601,PhysRevB.99.024404}. 
Otherwise, response functions diverge at zero temperature, which is an unphysical behavior.

In this paper, we propose a proper calculation for the \textit{orbital} gravito-ME response within a full quantum mechanical calculation. 
First, we show  in Sec.~\ref{formalization} that the second derivative of the current energy-density correlation function includes the information on the orbital gravito-ME tensor.
Second, we show in Sec.~\ref{OMQM_correction} that the intrinsic orbital gravito-ME response needs a correction from a higher-order magnetic multipole moment, an orbital magnetic quadrupole moment. Third, we derive equations for the intrinsic and extrinsic orbital gravtio-ME effects in periodic metals at finite temperature in Sec.~\ref{set_up}, Sec.~\ref{iOGME}, and Sec.~\ref{eOGME}. In addition, we show that these equations satisfy the Mott relation, similar to the relationship between electrical conductivity and thermoelectric conductivity, and that the intrinsic part has no unphysical divergence at zero temperature. Fourth, we show in Sec.~\ref{symmetry} the classification table for these responses with the magnetic point group and show that the intrinsic part can be used for the detection of $\mathcal{PT}$-symmetric orders. Finally, we calculate the intrinsic part in a model of a $\mathcal{PT}$-symmetric orbital magnet with a loop current order in Sec.~\ref{model_calculation} and demonstrate the enhancement around Dirac points.

\section{Formalization} \label{formalization}
In this section, we discuss the formalization of the orbital magnetization linearly induced by a temperature gradient, which we call the orbital gravito-magnetoelectric (OGME) response in this paper.
In general, it is challenging to calculate the orbital magnetization using the Bloch basis because the position operator is unbounded. However, the orbital ME tensor is known to be contained in the current-current correlation function and the current-density correlation function, providing a way to calculate the orbital ME tensor without using the position operator \cite{PhysRevB.82.245118,PhysRevLett.116.077201,shinada2022}. Here, we will see that the OGME tensor can also be extracted from a correlation function, i.e., the linear response function of the current density induced by a temperature gradient.

Before presenting the formalization, we briefly comment on how to calculate correlation functions induced by a temperature gradient. We need to consider the temperature gradient as an external force; however, this force cannot be simply added to the microscopic Hamiltonian because it is a statistical force. J. Luttinger derived a method to treat the temperature gradient as a mechanical force by introducing a gravitational potential $\psi(\x)$ \cite{PhysRev.135.A1505}. He 
showed that the response function induced by the gradient of the gravitational potential reproduces the temperature gradient-induced response function.
This gravitational potential is added to the unperturbed Hamiltonian $H_0$ as
\begin{eqnarray}
H_{\mathrm{grav}} = \int d\x \psi(\x) H_0(\x),
\end{eqnarray}
where $H_0(\x) = \{H_0 , \delta(\x - \bm{r}) \}/2$ is a Hamiltonian density. In the dynamical linear response theory, the total current density is changed by the gravitational potential with a factor $e^{-i\omega t + \delta t}$ as
\begin{eqnarray}
J^i_{\q,\omega} = \Phi^i(\q,\omega) \psi_{\q}.
\label{current}
\end{eqnarray}
$\Phi^i(\q,\omega)$ is the dynamical linear response function of the current density induced by the gravitational potential, which is discussed in this paper. This correlation function can be calculated by the usual Kubo formula because the gravitational potential is just a mechanical force.

Next, we will move on to the formalization.
We will show that the second derivative of this correlation function $\Phi^{i,jk}(\omega) = \partial_{q_j q_k} \Phi^i(0,\omega)$ includes the information of the OGME tensor.
$\Phi^{i,jk}(\omega)$  is symmetric for the interchange $j \leftrightarrow k$. Thus, we can decompose this function using a traceless rank-2 tensor, $\beta_{ij}$, and a totally symmetric rank-3 tensor $\gamma_{ijk}$, as
\begin{subequations}
\begin{align}
& \Phi^{k,ij}(\omega)
=
i \varepsilon_{jkl} \beta_{il}(\omega) + i \varepsilon_{ikl} \beta_{jl}(\omega) + \omega \gamma_{ijk}(\omega) \label{jh_correlation} \\
& \beta_{li}(\omega) = \frac{1}{3i} \varepsilon_{ijk} \Phi^{k,lj}(\omega) \label{beta} \\
& \gamma_{ijk}(\omega) = \frac{1}{3\omega} \Bigl( \Phi^{i,jk}(\omega)+\Phi^{j,ki}(\omega)+\Phi^{ki,j}(\omega) \Bigr) .
\end{align}
\end{subequations}
Substituting Eq.~(\ref{jh_correlation}) into Eq.~(\ref{current}), we find
\begin{subequations}
\begin{align}
& J^i_{\q,\omega} = -i\omega \Bigl( -iq_j Q^{\bm{G}ij}_{\q,\omega} \Bigr) + \Bigl( i\q \times \bm{M}^{\bm{G}}_{\q,\omega} \Bigr )_i \\
& Q^{\bm{G}ij}_{\q,\omega} = - \gamma_{ijk}(\omega) G^k_{\q,\omega} \\
& M^{\bm{G}i}_{\q,\omega} = 2i \beta_{ji}(\omega) G^j_{\q,\omega},
\end{align}
\end{subequations}
where $\bm{G}_{\q,\omega} = -i\q \psi_{\q,\omega}$ corresponds to a temperature gradient. 
$\bm{M}^{\bm{G}}_{\q,\omega}$ and $Q^{\bm{G}ij}_{\q,\omega}$ are a magnetization and an electric quadrupole moment induced by the temperature gradient $\bm{G}_{\q,\omega}$. Thus, $\beta_{ij}$ can be interpreted as the OGME tensor, and $\gamma_{ijk}$ is a pure electric quadrupole moment induced by the temperature gradient, which cannot be included in the OGME tensor due to its total symmetric property. In this paper, we will derive the static and uniform OGME response by calculating $\beta_{ij}$.

\section{Derivation of the OGME tensor} \label{derve_of_OGME}
In this section, we give the result of the static and uniform OGME tensor in periodic systems at finite temperatures. 
When we take the static and uniform limits, we take $\q \to 0$ before $\omega \to 0$ because we focus on the dynamical response.
\subsection{Orbital magnetic quadrupole moment correction} \label{OMQM_correction}
When we consider the response by the gravitational force, we should note that the current density is also changed by the gravitational potential. The total current density under this potential is given by
\begin{eqnarray}
\J(\x) = (1 + \psi(\x)) \J_0(\x), \label{current_psi}
\end{eqnarray}
where $\J_0(\x)$ is the unperturbed current density. Thus, there are two contributions to the linear response function induced by the gravitational force. One contribution comes from the first term in Eq.~(\ref{current_psi}). This contribution originates from the change of the density matrix by the gravitational force, and it is given by the usual Kubo linear response.
It brings the correlation function between the current density and the Hamiltonian density, which we call the current energy-density correlation function $\Phi^i_{JH}(\q,\omega)$. 
Another contribution originates from the second term. This term is already linear in the gravitational potential. Thus, we need to calculate the equilibrium expectation value of the current density $\J_0(\x)$. This equilibrium current plays an important role in thermal responses. In fact, it is known to give an orbital (energy) magnetization correction to the Nernst conductivity and the thermal Hall conductivity, and it eliminates an unphysical divergence at zero temperature \cite{LSmrcka_1977,PhysRevLett.107.236601} and these conductivities satisfy the Mott relation.
We will see that the equilibrium current adds a higher-order multipole correction, an orbital magnetic quadrupole moment, to the OGME tensor, in an analogy to the thermal Hall effect.

Let us discuss the contribution from the equilibrium current in more detail.
In equilibrium, a bulk current is not allowed to flow. However, a local current is not forbidden. In general, a local current can be written as a local magnetization $\bm{M}(\x)$ using $\Braket{\J_0(\x)}_0 = \bm{\nabla} \times \bm{M}(\x)$, which is called a magnetization current.
Here, $\braket{O}_0$ is an expectation value of an observable operator $O$ in equilibrium.
This magnetization can be derived in a thermodynamic approach \cite{PhysRevLett.99.197202}. In local thermodynamics, the free energy density $F(\x)$ with a nonuniform magnetic field $\bm{B}(\x)$ is given by
\begin{eqnarray}
dF(\x) = M_0^i(\x) d B^i(\x) + Q^{ij}(\x) d[\partial_i B^j(\x)] + \cdots .
\end{eqnarray}
This thermodynamic relation defines multipoles conjugate to the higher-order gradient of the magnetic field, where $Q^{ij}(\x)$ is a local magnetic quadrupole moment. Using this relationship, the local magnetization $\bm{M}(\x)$ is determined by
\begin{eqnarray}
M^i(\x) &=& \partial F(\x) / \partial B^i(\x) \nonumber \\
&=& M^i_0(\x) - \partial_j Q^{ji}(\x) + \mathcal{O}(\partial^2). 
\end{eqnarray}
The bulk magnetization and magnetic quadrupole moment in periodic systems can be calculated using this thermodynamic definition, where their formulas are applicable even in metals at finite temperatures \cite{PhysRevLett.99.197202,PhysRevB.98.020407}.

Using the local magnetization, the contribution from the second term in Eq.~(\ref{current_psi}) can be rewritten as
\begin{eqnarray}
\psi(\x) \Braket{\J_0(\x)}_0 &=& \psi(\x) \bm{\nabla} \times \bm{M}(\x) \nonumber \\
&=&
\bm{\nabla} \times \{ \psi(\x) \bm{M}_0(\x) - \partial_i (\bm{e}_j \psi(\x) Q_{ij}(\x) )   \} \nonumber \\
&&
+ \bm{\nabla} \times \{  (\partial_i \psi(\x)) \bm{e}_j Q_{ij}(\x) \} \nonumber \\
&&
-\bm{\nabla} \psi(\x) \times (\bm{M}_0(\x) - \partial_i\bm{e}_jQ_{ij}(\x)).
\end{eqnarray}
Here, $\bm{e}_i$ is a unit vector along the $i$-direction.
The first term is the gravitational potential correction to the magnetization and the magnetic quadrupole moment. The third term represents the current-density response induced by a temperature gradient where the conductivity is determined by the local magnetization. This contribution is studied in the Nernst conductivity \cite{LSmrcka_1977,PhysRevLett.107.236601}. The second term is the main result of this paper. This term corresponds to a magnetization induced by a temperature gradient, i.e., the gravito-ME effect. This means that the magnetic quadrupole moment should be included in the OGME tensor besides the contribution from the current energy-density correlation function as $\chi^{\mathrm{OGME}}_{ij} = 2i\beta^{JH}_{ij} + Q_{ij}$ ($M^i = \chi^{\mathrm{OGME}}_{ji} (\partial_j \psi)$).

The quadrupole moment correction is also studied in the \textit{spin} gravito-ME effect, where the \textit{spin} magnetic quadrupole moment contributes to the response tensor \cite{PhysRevB.99.024404}. With the gravitational potential, the spin density also changes as $(1+\psi(\x)) \bm{s}(\x)$, where the second term gives the contribution from the spin magnetic quadrupole moment. Our equation, on the other hand, includes the \textit{orbital} quadrupole moment.

\subsection{Set-up} \label{set_up}
We have shown above that we need to calculate the current energy-density correlation function and the magnetic quadrupole moment to obtain the OGME tensor. In the following, we will derive the OGME tensor in periodic systems described by a Bloch Hamiltonian. 

First, we will derive the current energy-density correlation function.
The Hamiltonian used in this paper is 
\begin{eqnarray}
    H_0 = \frac{\bm{p}^2}{2m} + V(\bm{x}) + \frac{1}{4m^2} \biggl( \frac{\partial V(\x)}{\partial \x} \times \bm{p} \biggr) \cdot \bm{\sigma}.
\end{eqnarray}
Here, $m$ is the electron mass, $V(\x)$ is a periodic potential, and $\bm{\sigma}$ is the vector of the Pauli matrices. The third term represents the spin-orbit coupling. The unperturbed current operator is $\J_0(\bm{r},t) = -e \{\bm{v}_0 ,\delta(\x- \bm{r}) \}$ and $\bm{v}_0 = i[H_0 , \bm{x}]$ is the velocity operator.
We note that $\bm{r}$ is here just a coordinate and not an operator, unlike $\x$.
Using the Kubo formula, the current energy-density correlation function is given by
\begin{widetext}
\begin{eqnarray}
\Phi^{i}_{JH}(\q,\omega) 
= 
-e \sum_{mn,\kvec} \frac{f(\epsilon_{n\kvec+\q/2}) - f(\epsilon_{m\kvec-\q/2})}{\epsilon_{n\kvec+\q/2} - \epsilon_{m\kvec-\q/2} -( \omega +  i\delta )} 
\bra{u_{m\kvec-\q/2}} v^i_{\kvec} \ket{u_{n\kvec+\q/2}} \frac{\epsilon_{n\kvec+\q/2} + \epsilon_{m\kvec-\q/2}}{2} \braket{u_{n\kvec+\q/2} | u_{m\kvec-\q/2}} . \nonumber \\
\end{eqnarray}
We define $\epsilon_{n\kvec}$, $\ket{u_{n\kvec}}$ as an eigenenergy and eigenvector of the $n$-th band of the Bloch Hamiltonian $H_{\kvec}=e^{-i\kvec \cdot \x} (H_0 -\mu N) e^{+i\kvec \cdot \x}$. $f(\epsilon) = 1/(e^{\beta \epsilon} + 1)$ is the Fermi distribution function, and $\bm{v}_{\kvec} = \partial H_{\kvec}/\partial \kvec$ is the velocity operator.

\subsection{Intrinsic OGME tensor} \label{iOGME}
After having defined the current energy-density correlation function in periodic systems, we expand the correlation function up to the second-order of $\q$ and use the relation in Eq.~(\ref{beta}) to derive $\beta^{JH}_{ij}(0)$, as a contribution to the OGME tensor originating from the correlation function. 
In this subsection, we focus on the intrinsic part. 
The result for the extrinsic part will be shown and discussed in the next subsection.
Details of the derivation are given in Appendix \ref{detail_current-enegy}.
The intrinsic part from the correlation function ($2i \beta^{JH}_{li} = \frac{2}{3} \varepsilon_{ijk} \Phi^{k,lj}_{JH}$) is
\begin{eqnarray}
2i \beta^{JH}_{li}
&=&
-e \int_{\mathrm{BZ}} \frac{d^3k}{(2\pi)^3} \sum_{n} \biggl[
\epsilon_{n\kvec} f(\epsilon_{n\kvec}) \Bigl\{
\frac{1}{3} \varepsilon_{ijk} \partial_k g^{lj}_n -  \sum_{m(\neq n)}\frac{2}{\epsilon_{nm\kvec}} \R [\A^l_{nm} M^i_{mn}] \Bigr\}
\nonumber \\
&& \hspace{80pt}
+ f(\epsilon_{n\kvec})
\Bigl\{
\frac{2}{3} \sum_{m(\neq n)} \R [\A^l_{nm} M^i_{mn}] 
- \frac{1}{12} \varepsilon_{ijk} \partial_j v^{kl}_n
\Bigr\}
\biggr]. \label{beta_JH} 
\end{eqnarray}
We use the notation $\partial_i  = \partial  /\partial k_i$, $\epsilon_{nm\kvec} = \epsilon_{n\kvec} - \epsilon_{m\kvec}$, and $v^{ij}_n = \bra{u_{n\kvec}} \partial^2 H_{\kvec}/ \partial k_i \partial k_j \ket{u_{n\kvec}}$. This equation includes several geometric quantities. First, $\A^i_{nm} = i\braket{u_{n\kvec} | \partial_i u_{m\kvec}}$ is the Berry connection. Second, $g^{ij}_n = \sum_{m(\neq n)} \R[\A^i_{nm} \A^j_{mn}]$ is the quantum metric measuring the distance of two states on the Brillouin zone \cite{provost1980riemannian,resta2011insulating}. It has been shown recently that the quantum metric appears in several quantities, such as the electric quadrupole moment \cite{PhysRevB.102.235149}, the superfluid weight \cite{peotta2015superfluidity}, the nonlinear response \cite{PhysRevLett.112.166601,PhysRevX.11.011001}, the nonreciprocal directional dichroism \cite{PhysRevLett.122.227402}, and the orbital ME effect \cite{PhysRevB.103.045401,PhysRevB.103.115432,shinada2022}. Third, $\bm{M}_{nm} = \sum_{l(\neq n)}\bm{V}_{ml,n} \times \bm{\A}_{ln}$ ($\bm{V}_{ml,n} = (\bm{v}_{\kvec}^{ml} + \bm{\nabla} \epsilon_{n\kvec} \delta_{ml})/2$) corresponds to the product of the velocity and the position; thus it corresponds to an off-diagonal orbital magnetic moment.

The orbital magnetic quadrupole moment in periodic systems has already been derived using the thermodynamic approach \cite{PhysRevB.98.020407}. We reproduce it using our model and obtain
\begin{eqnarray}
Q_{li}
&=&
-e \int_{\mathrm{BZ}} \frac{d^3 k}{(2\pi)^3} \sum_{n}
\biggl[
-g(\epsilon_{n\kvec}) \Bigl\{
\frac{1}{3} \varepsilon_{ijk} \partial_k g^{lj}_n -  \sum_{m(\neq n)}\frac{2}{\epsilon_{nm\kvec}} \R [\A^l_{nm} M^i_{mn}]
\Bigr\} \nonumber \\
&& \hspace{80pt}
-
f(\epsilon_{n\kvec}) 
\Bigl\{
\frac{2}{3} \sum_{m(\neq n)} \R [\A^l_{nm} M^i_{mn}]
-
\frac{1}{12} \varepsilon_{ijk} \partial_j v^{kl}_n
\Bigr\}
\biggr]. \label{m_quad}
\end{eqnarray}
Here, $g(\epsilon) = -\beta^{-1} \log ( 1+ e^{-\beta \epsilon})$ is the grand potential density. This equation is identical to the equation in Ref.~\cite{PhysRevB.98.020407,PhysRevB.98.060402}. 

Finally, combining Eq.~(\ref{beta_JH}) and Eq.~(\ref{m_quad}), the actual total intrinsic OGME tensor $\chi^{\mathrm{iOGME}}_{ij}$ ($M^i = \chi^{\mathrm{iOGME}}_{ji} \partial_j T $) using the correspondence $\partial_i \psi \to - \partial_i T /T$ is
\begin{eqnarray}
\chi^{\mathrm{iOGME}}_{ij}
&=&
-(2i\beta^{JH}_{ij} + Q_{ij})/T
=
e\int_{\mathrm{BZ}} \frac{d^3k}{(2\pi)^3}
\sum_{n} s(\epsilon_{n\kvec}) \biggl(
\frac{1}{3} \varepsilon_{klj} \partial_{l} g^{ik}_{n\kvec} - \sum_{m(\neq n)} \frac{2}{\epsilon_{nm\kvec}}
\R \Bigl[ 
\A^i_{nm} M^j_{mn}
\Bigr]
\biggr). \label{intrinsicOGME}
\end{eqnarray}
\end{widetext}
In this calculation, we can see that terms proportional to the Fermi distribution function in Eq.~(\ref{beta_JH}) and Eq.~(\ref{m_quad}) cancel each other.
This equation is the main result of this paper. $s(\epsilon) = \epsilon f(\epsilon)/T - g(\epsilon)/T $ is an entropy density. The OGME tensor is very similar to the orbital ME tensor (see Ref.~\cite{shinada2022}). The only difference is that the entropy density $s(\epsilon_{n\kvec})$ in the OGME tensor is replaced by the Fermi distribution function $f(\epsilon_{n\kvec})$ in the orbital ME tensor. It is natural that the entropy density appears in the thermal response. Because the entropy density becomes zero at zero temperature, the OGME tensor approaches zero with decreasing temperature and has no unphysical divergence, which used to be a problem for the thermal Hall effect. In addition, the similarity to the  orbital ME tensor leads to the following relation 
(see Appendix \ref{derivation_Mott} for the details)
\begin{eqnarray}
\chi^{\mathrm{iOGME}}_{ij} (\mu,T) 
&=& 
\int d\epsilon \frac{(\epsilon - \mu)}{eT} \frac{\partial f(\epsilon-\mu)}{\partial \epsilon} \chi^{\mathrm{iOME}}_{ij}(\epsilon,T=0) \nonumber \\
&\simeq&
\frac{-\pi^2 T}{3e} \frac{\partial \chi^{\mathrm{OME}}_{ij}(\mu,0)}{\partial \mu} \hspace{10pt} (T \to 0) \label{Mott}
\end{eqnarray}
Here, $\chi^{\mathrm{iOME}}_{ij}(\mu,T=0)$ is the intrinsic orbital ME tensor at the chemical potential $\mu$ and zero temperature.
This equation is known as the Mott relation, which was first introduced as a relation between electric conductivity and thermoelectric conductivity \cite{mott1936theory}.
Now, this relation is known to hold for various responses, such as the quantum anomalous Hall current \cite{PhysRevLett.97.026603} and the spin ME effect \cite{PhysRevB.99.024404}.
Thus, a similar relation is valid for the orbital magnetization induced by an electric field and a temperature gradient. The second line of this equation shows the behavior at low temperatures demonstrating that the OGME tensor scales $T$-linear at low temperatures.

We note that the trace of our equation is zero. The trace includes the topological term that is called the Chern-Simons
term. 
In addition, we note that our formula is gauge invariant because it is only written by the off-diagonal Berry connection, which is gauge invariant.
Finally, we comment on previous works. The intrinsic OGME tensor has been studied using the semiclassical theory in Ref.~\cite{PhysRevB.103.045401,PhysRevB.103.115432}. There is a small difference to our formula, i.e., the $1/3$ factor in Eq.~(\ref{intrinsicOGME}) is replaced by $1/2$ in Ref.~\cite{PhysRevB.103.045401,PhysRevB.103.115432}.

\begin{table*}[t]
\caption{Table classifying whether the extrinsic part and the intrinsic part of the (gravito-) magnetoelectric response can be finite in a magnetic point group (MPG). The results are obtained using MTENSOR in the Bilbao Crystallographic Server \cite{aroyo2006bilbaoI,aroyo2006bilbaoII,aroyo2011crystallography}. \checkmark corresponds to a finite response, while - corresponds to a vanishing response. Groups not included in this table do not produce either response. The symbols $\mathcal{P}$ and $\mathcal{T}$ correspond to the inversion symmetry and the time-reversal symmetry. $\mathcal{PT}$ represents the operation of the product of $\mathcal{P}$ and $\mathcal{T}$.}
\label{table_MPG}
\begin{tabular}{ll|cc}
&MPG & intrinsic & extrinsic \\
\hline \hline
&&& \\
($\mathcal{PT}\bigcirc$)
&$\bar{1}',2'/m,2/m',m'mm,m'm'm',4/m',4'/m',4/m'mm,4'/m'm'm,4/m'm'm'$   & \checkmark & - \\
&$\bar{3}',\bar{3}'m,\bar{3}'m',6/m',6/m'mm,6/m'm'm'$ &&\\
($\mathcal{PT}\times$)
& $\bar{6}',\bar{6}'m'2,\bar{6}'m2'$ &&\\
&&& \\
\hline
&&& \\
 &$1,2,2',m,m',222,2'2'2,mm2,m'm2',m'm'2,4,4',\bar{4},\bar{4}'$ & \checkmark & \checkmark \\
 &$422,4'22',42'2',4mm,4'm'm,4m'm',\bar{4}2m,\bar{4}'2'm,\bar{4}'2m',\bar{4}2'm'$ && \\
 &$3,32,32',3m,3m',6,622,62'2',6mm,6m'm'$  & & \\
&&& \\
\hline
&&& \\
($\mathcal{T}\bigcirc$)
&$11',21',m1',2221',mm21',41',\bar{4}1',4221',4mm1',\bar{4}2m1',31',321',3m1'$
 &  - & \checkmark \\
&$61',6221',6mm1',[231',4321']$  &&\\
($\mathcal{T}\times$)
&$6',6'22',6'mm',[23,432,4'32']$    &&\\
\end{tabular}
\end{table*}

\subsection{Extrinsic OGME tensor} \label{eOGME}
In this subsection, we will discuss the extrinsic part of the OGME tensor. Using the relation in Eq.~(\ref{beta}), the extrinsic part $\chi^{\mathrm{eOGME}}_{ij}$ ($M^i = \chi^{\mathrm{eOGME}}_{ji} (\partial_j T)$) is (see Appendix \ref{detail_current-enegy} for a detailed derivation)
\begin{eqnarray}
\chi^{\mathrm{eOGME}}_{ij} &=& -2i \beta^{JH}_{ij}/T \nonumber \\
&=&
\frac{e}{\delta T} \int_{\mathrm{BZ}} \frac{d^3k}{(2 \pi)^3} \sum_{n} \epsilon_{n\kvec} \frac{\partial f(\epsilon_{n\kvec})}{\partial k_i} m^j_{n\kvec}. \nonumber \\ 
\end{eqnarray}
Here, $\bm{m}_{n\kvec} = \I [\bra{ \bm{\nabla} u_{n\kvec}} \times (\epsilon_{n\kvec} - H_{\kvec})  \ket{\bm{\nabla} u_{n\kvec}}]/2$ is the orbital magnetic moment (OMM), and $\delta$ is an adiabatic factor corresponding to the inverse of the dissipation strength. Due to $\delta$, the extrinsic part has a Drude-like singularity. Furthermore, it originates at the Fermi surface.

Here, we shortly comment on the time-reversal symmetry in the case of the extrinsic OGME tensor. The integrand in this equation is even against the time-reversal operation because the OMM $\bm{m}_{n\kvec}$ and the Bloch wave vector $\kvec$ are odd. As mentioned above, the (gravito-) ME effect needs time-reversal symmetry breaking. In the case of the extrinsic OGME, the change of the distribution function, together with dissipation, effectively breaks  time-reversal symmetry. Thus, the Hamiltonian of the system does not need to break time-reversal symmetry. For this reason, the extrinsic part is better interpreted as a current inducing the magnetization ($M^i = \chi_{ji} J^j$), which is usually called the Edelstein effect.

The extrinsic part also satisfies the Mott relation (see Appendix \ref{derivation_Mott} for details)
\begin{eqnarray}
\chi^{\mathrm{eOGME}}_{ij} (\mu,T) 
&=& 
\int d\epsilon \frac{(\epsilon - \mu)}{eT} \frac{\partial f(\epsilon-\mu)}{\partial \epsilon} \chi^{\mathrm{eOME}}_{ij}(\epsilon,T=0) \nonumber \\
&\simeq&
\frac{-\pi^2 T}{3e} \frac{\partial \chi^{\mathrm{eOME}}_{ij}(\mu,0)}{\partial \mu} \hspace{10pt} (T \to 0). \label{eMott}
\end{eqnarray}
Here, $\chi^{\mathrm{eOME}}_{ij}(\epsilon,T=0)$ is the extrinsic orbital ME tensor at the chemical potential $\mu$ and zero temperature.

\section{symmetry analysis and model calculation} \label{symmwtry_model}

\subsection{Magnetic point group symmetry analysis} \label{symmetry}
In the above discussion, we have shown that there are two parts to the OGME response, an extrinsic part and an intrinsic part. In general, the breaking of both the time-reversal symmetry and the inversion symmetry is needed for the (gravito-) ME response. Thus, we need to use magnetic point groups (MPG) to analyze whether the responses can exist. 
As explained above, the extrinsic part is a dissipation effect that does not need a time-reversal symmetry breaking of the Hamiltonian. 
Due to the difference in how both parts change with respect to the time-reversal symmetry, the conditions for their appearance are different. In fact, the extrinsic part is an axial and time-reversal-even rank-2 tensor, and the intrinsic part is an axial and time-reversal-odd rank-2 tensor. The classification table is created according to their symmetries and shown in Table~\ref{table_MPG}. 

Let us make some comments on the table.
When the system obeys $\mathcal{PT}$-symmetry, the product of inversion symmetry $\mathcal{P}$ and time-reversal symmetry $\mathcal{T}$, the orbital magnetic moment $\bm{m}_{n\kvec}$ vanishes. Thus, the extrinsic part cannot appear. Most of the symmetry groups generating only the intrinsic part obey $\mathcal{PT}$-symmetry; however, three groups without $\mathcal{PT}$-symmetry, ($\bar{6}',\bar{6}'m'2,\bar{6}'m'2'$), 
also only have an intrinsic response, as shown in Table~\ref{table_MPG}.
There are $21$ $\mathcal{PT}$-symmetric groups out of $122$ magnetic point groups. Thus, Table~\ref{table_MPG} shows that the intrinsic response is sensitive for detecting many of the $\mathcal{PT}$-symmetric MPGs. 
On the other hand, when the system fulfills $\mathcal{T}$ symmetry, the intrinsic part is forbidden. 

Let us now comment on the classification of the intrinsic part. We show in Table~\ref{table_MPG} that there are $53$ groups with a finite intrinsic part. However, there are $58$ groups allowing for the intrinsic ME effect in general. As mentioned above, our equation of the intrinsic part (Eq.~(\ref{intrinsicOGME})) is traceless. In other words, the monopole term is zero. Thus, the intrinsic part vanishes in five groups ($23,m'\bar{3}',432,\bar{4}'3m',m'\bar{3}'m'$). The monopole term, such as the Chern-Simons term, is allowed for these five groups.

Moreover, we comment on the relation with the nonlinear (second-order) Hall effect. In general, the nonlinear Hall effect consists of an extrinsic Hall effect induced by the Berry curvature dipole \cite{PhysRevLett.115.216806} and an intrinsic Hall effect \cite{PhysRevLett.112.166601,PhysRevX.11.011001}. The extrinsic part is written by an axial and time-reversal-even rank-2 tensor, and the intrinsic part is written by an axial and time-reversal-odd rank-2 tensor. Thus, the conditions for the appearance are identical to the extrinsic part and the intrinsic part of the orbital (gravito-) ME effect \cite{PhysRevLett.127.277201}. However, there is a small difference between the extrinsic ME effect and the nonlinear Hall effect.
The extrinsic nonlinear Hall effect is traceless due to $\bm{\nabla} \cdot \bm{\Omega}_{n\kvec} = 0$. Thus, 
the extrinsic nonlinear Hall effect vanishes for the 
groups in the square bracket ([]) in Table~\ref{table_MPG}. However, the extrinsic ME effect can appear because the orbital magnetic moment $\bm{\nabla} \cdot \bm{m}_{n\kvec}$ is not necessarily zero. This classification shows that the orbital (gravito-) ME effect is useful as another detection method in addition to the nonlinear Hall effect. Especially, the nonlinear Hall effect is zero in insulators; thus, the ME effect seems better suited in insulators.

\begin{figure*}[t]
\includegraphics[width=0.45\linewidth]{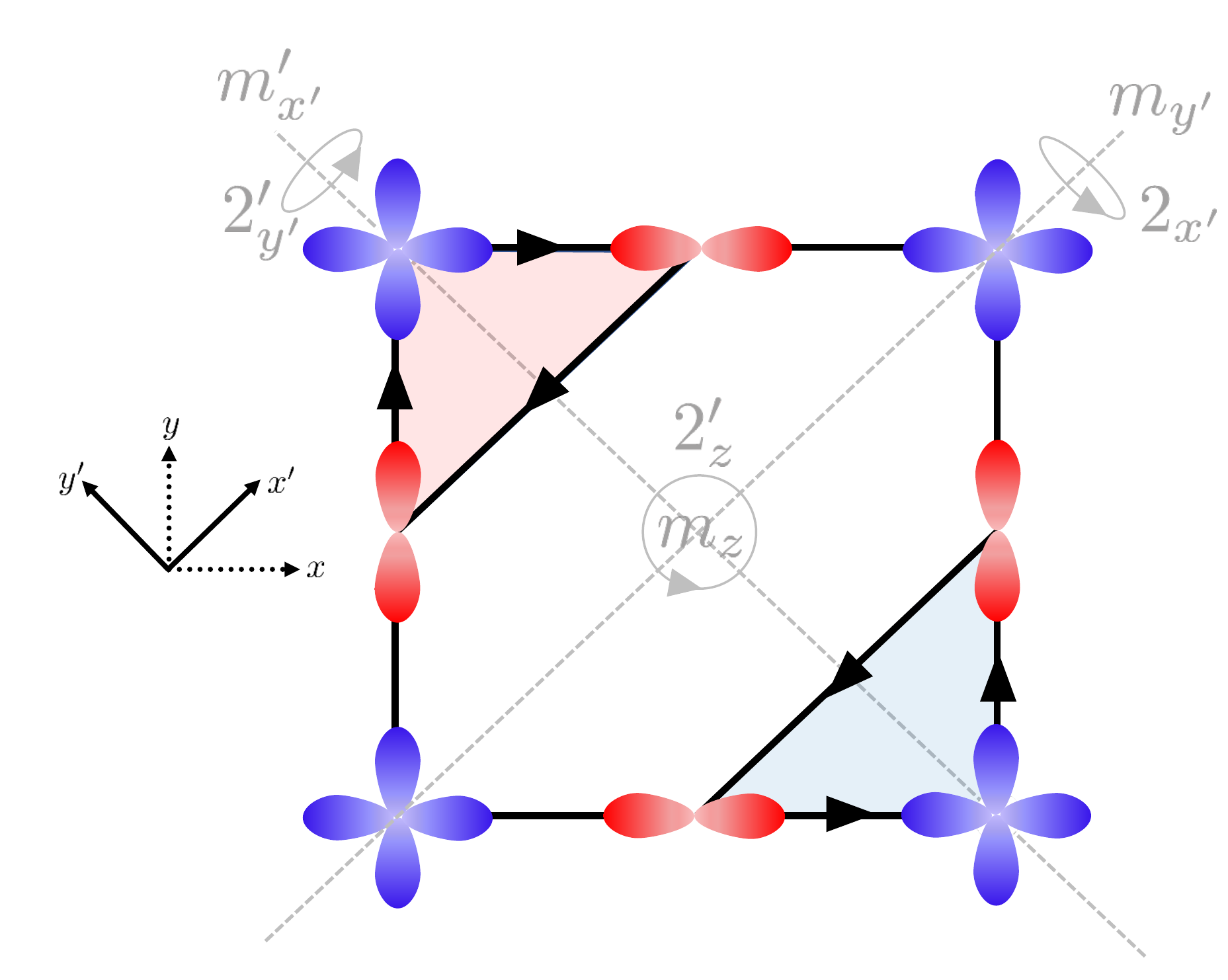}
\includegraphics[width=0.45\linewidth]{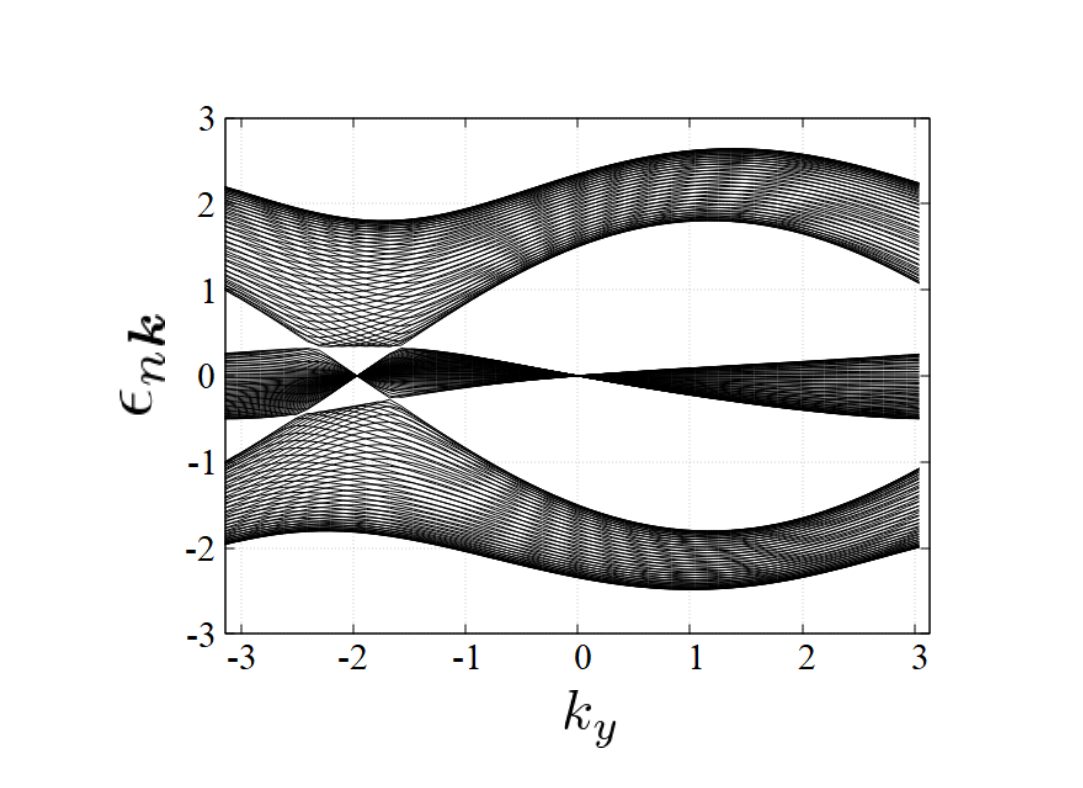}
\caption{(Left) Schematic diagram of the loop current in the $\mathrm{CuO_2}$ plane. The blue four-leaves and red two-leaves represent the $d$-orbitals on the copper sites and $p_x$, $p_y$-orbitals on the oxygen sites, respectively. The arrows represent the complex hopping characterizing loop currents. In this case, two $z$-directional orbital magnetic moments with opposite signs are shown by red and blue shaded areas. The gray letters represent the invariant symmetries in $m'mm$. $m_i$ is the mirror symmetry against the plane vertical to the $i$-axis and $2_i$ is the $180$ degrees rotational symmetry around the $i$-axis. The symbols with the prime ($'$), such as $2'_z$ represent the original symmetry multiplied by the time reversal operation $\mathcal{T}$. This model fulfills $\mathcal{PT}$-symmetry ($\mathcal{PT} = m'_{x'}m_{y'}m_z$). (Right) Band dispersion of our model. This model has four Dirac points at $E=-0.44t$, $-0.27t$, and $0.33t$.
}  \label{ybco}
\end{figure*}

\subsection{Model calculation} \label{model_calculation}
In this subsection, we calculate the intrinsic OGME tensor in a specific model. In the previous subsection, we have shown the list classifying allowed groups for the OGME tensor. Groups in the top row in Table \ref{table_MPG} are useful to detect the intrinsic OGME response because the extrinsic part vanishes. 

Furthermore, orbital magnetization does, in general, not need spin degrees of freedom. In other words, the orbital ME effect can be finite in systems with only an orbital component. Thus, the orbital ME effect will help detect orbital magnetic orders, and we focus here on a system exhibiting orbital magnetic order, i.e., the loop current order.

Loop currents have been mainly discussed in the pseudogap phase of the high-temperature superconductors \cite{PhysRevB.55.14554,PhysRevLett.96.197001,li2008unusual,PhysRevLett.111.047005,zhao2017global,BOURGES2011461,CRPHYS_2021__22_S5_7_0}. Recently, loop current orders are also discussed as candidates of the time-reversal symmetry breaking charge-order in the Kagome superconductors $\mathrm{AV_3Sb_5}$ (A=K, Rb, Cs) \cite{mielke2022time}, the hidden order in the spin-orbit coupled Mott insulator $\mathrm{Sr_2IrO_4}$ \cite{PhysRevX.11.011021,zhao2016evidence,jeong2017time}, and the orbital ferromagnetic phase with a loop current in twisted bilayer graphenes \cite{PhysRevX.9.031021,PhysRevB.103.035427,PhysRevX.10.031034,liu2021orbital}. Here, we will study the loop current order in cuprates. Several proposals of loop current orders exist for this class. We focus here on the $\mathcal{PT}$-symmetric order shown in Fig.~\ref{ybco}, where two opposite local currents (red shaded area and blue one) occur. This order is called LC-$\Theta_{\mathrm{II}}$ state \cite{PhysRevB.55.14554} and belongs to $m'mm$ in the MPG. Thus, only the intrinsic part exists. The model Hamiltonian belonging to $m'mm$ is given by \cite{PhysRevB.98.060402,BOURGES2011461,PhysRevB.85.155106,shinada2022}
\begin{eqnarray}
H_{\kvec} = 
\begin{pmatrix}
0 & its_x + ir c_x & its_y + irc_y \\
-its_x -irc_x & 0 & t' s_x s_y \\
-its_y -i rc_y & t' s_x s_y & 0 
\end{pmatrix} ,
\end{eqnarray}
where $s_i=\sin(k_i/2a)$ and $c_i=\cos(k_i/2a)$ ($a$ is the lattice constant). This Hamiltonian has no spin degrees of freedom because we focus only on the orbital order. The basis consists of the $d$-orbitals $\ket{d}$ on the copper sites and the $p$-orbitals $\ket{p_x}, \ket{p_y}$ on the oxygen sites. $t$ and $t'$ are hopping constants across these orbitals and $r$ is the order parameter of the loop current. This model has four Dirac points at $E = -0.44t$, $-0.27 t$, and $0.33 t$, shown in Fig.~\ref{ybco}.

Let us discuss the intrinsic OGME tensor for this model. To make the discussion easier, we transform the coordinates from $(k_x , k_y) \to (k_{x'} , k_{y'})$ as shown in Fig.~{\ref{ybco}}. 
First, we consider the constraints of symmetry.
The group $m'mm$ has the mirror symmetry $m_{y'}$, thus $\chi^{\mathrm{iOGME}}_{x'x'} = \chi^{\mathrm{iOGME}}_{x'z} = \chi^{\mathrm{iOGME}}_{y'y'} =
\chi^{\mathrm{iOGME}}_{zx'} =
\chi^{\mathrm{iOGME}}_{zz} = 0$. In addition, this group fulfills the product symmetry of mirror and  time-reversal symmetry $m'_{x'}$. Thus, $\chi^{\mathrm{iOGME}}_{x'y'} =
\chi^{\mathrm{iOGME}}_{y'x'} = 0$. Therefore, the remaining terms are $\chi^{\mathrm{iOGME}}_{y'z}$ and $\chi^{\mathrm{iOGME}}_{zy'}$.

\begin{figure*}[t]
\includegraphics[width=0.45\linewidth]{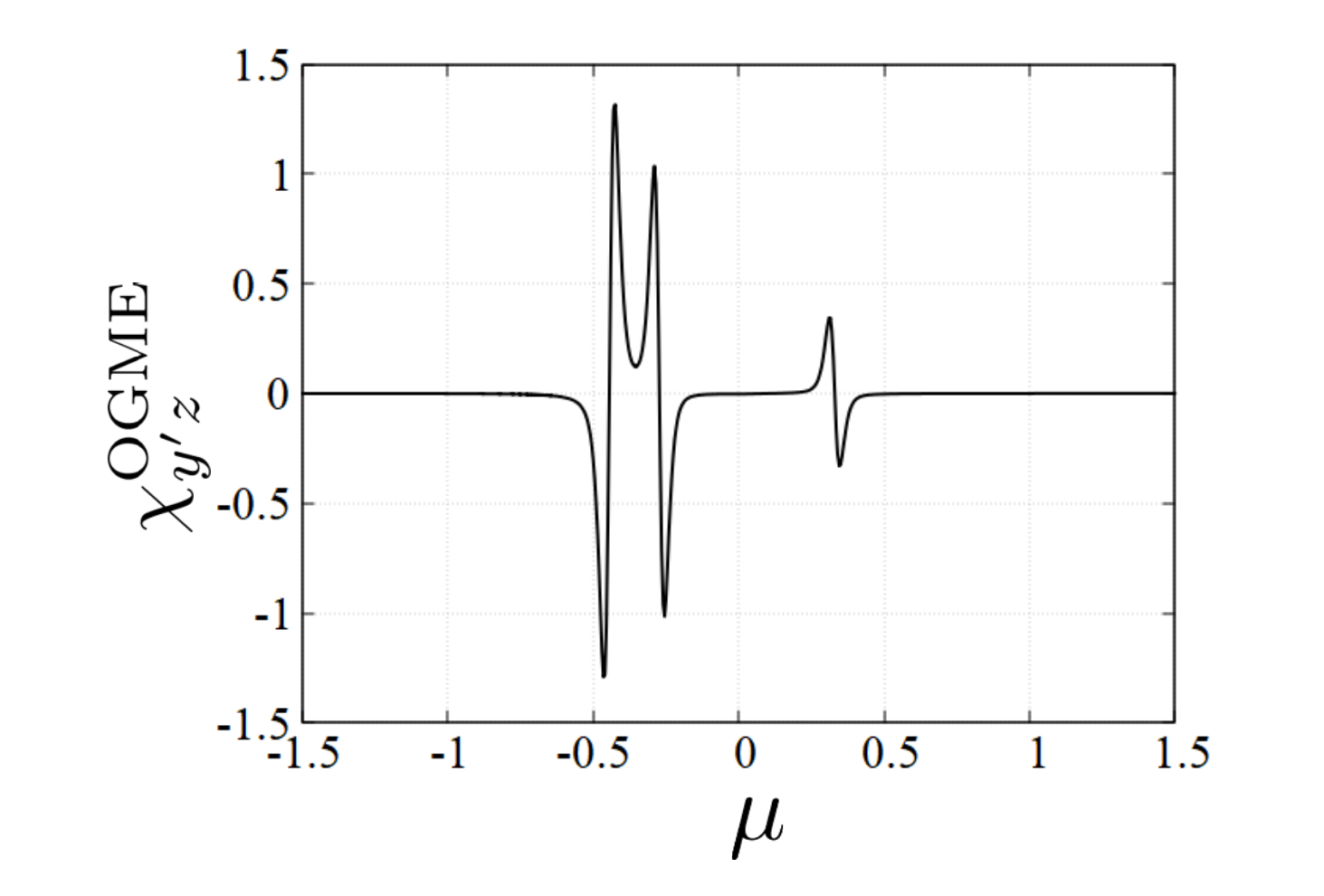}
\includegraphics[width=0.45\linewidth]{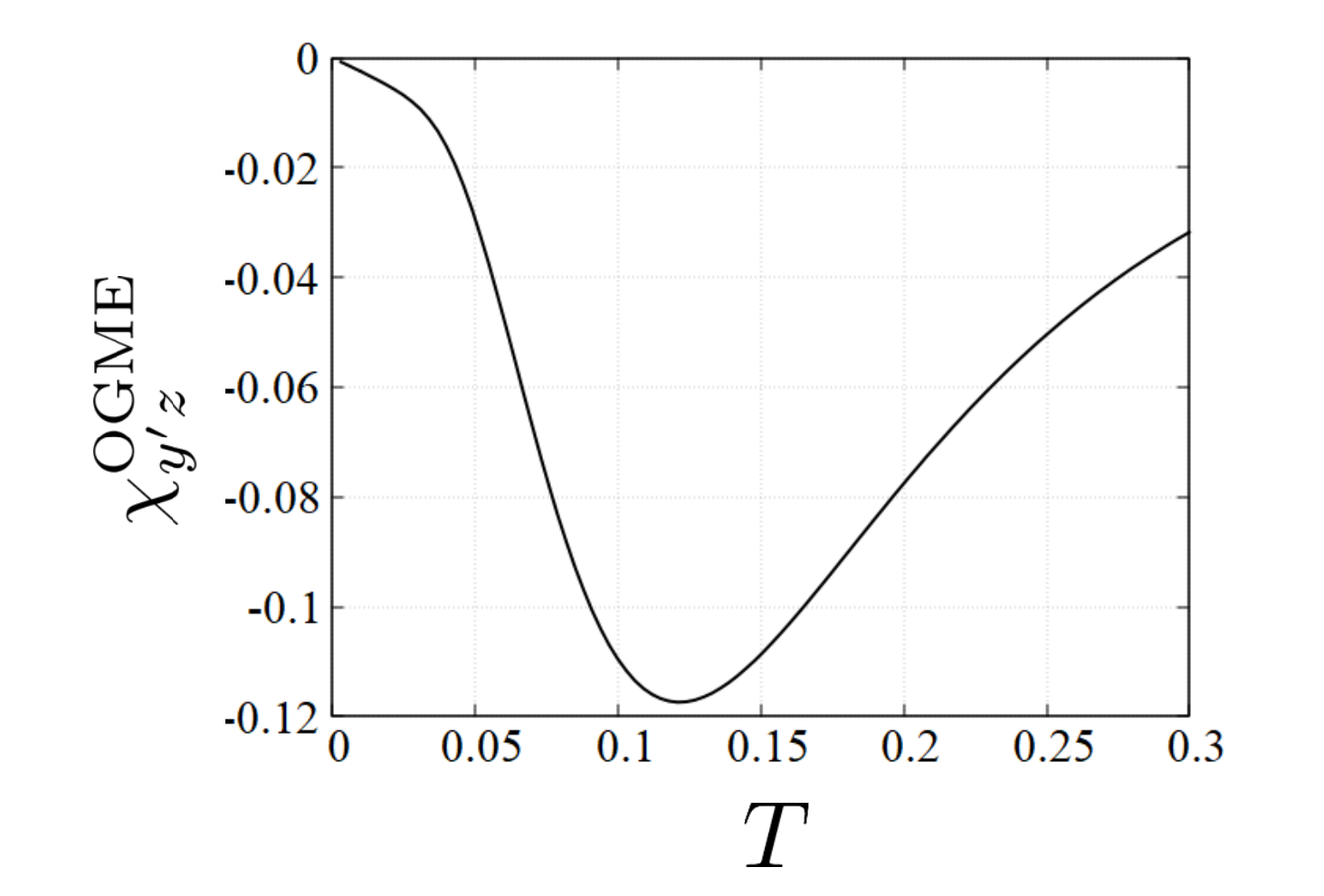}
\caption{(Left) The dependence of the OGME tensor on the chemical potential $\mu$ at $T/t=0.01$. (Right) The dependence of the OGME tensor on the temperature at $\mu = -0.8t$. We set $t=1.0$ and introduce an infinitesimal dissipation $\delta = 0.001i$ for the numerical calculation. We use $\chi^{\mathrm{OGME}}_{y'z}$ in units of $ea/\hbar$.}  \label{omt}
\end{figure*}

We calculate $\chi^{\mathrm{iOGME}}_{y'z}$ and show its dependence on the chemical potential $\mu$ and the temperature $T$ in Fig.~\ref{omt}. The left figure shows the dependence on the chemical potential at $T/t=0.01$, where we can see peak structures near the Dirac points. These structures also appear in the orbital ME response \cite{shinada2022}. As discussed above, the OGME tensor is determined by the orbital ME tensor (Eq.~(\ref{Mott})) and proportional to $\partial \chi^{\mathrm{iOGME}}_{ij} / \partial \mu$ at low temperature. This behavior can be confirmed when comparing this figure with Fig.~2 in Ref.~\cite{shinada2022}). Next, the right panel in Fig.~\ref{omt} shows the temperature dependence at $\mu = -0.8t$, where we can confirm the $T$-linear dependence at low temperature as derived in Eq.~(\ref{Mott}).

\section{Conclusions}
In conclusion, we have discussed the orbital magnetoelectric effect induced by a temperature gradient (orbital gravito-ME effect). We have derived its response using a full quantum approach.
We have shown that its response is formalized by the second derivative of the current energy-density correlation function. We have calculated the response in periodic metals, and we have shown the appearance of two terms, the extrinsic part and the intrinsic part. 
We have shown that the intrinsic part needs a correction from the orbital magnetic quadrupole moment.
Due to this correction, the intrinsic part has no unphysical divergence and satisfies the Mott relation. Furthermore, we have demonstrated that the extrinsic part also satisfies the Mott relation. In previous works, the intrinsic part was derived using a semiclassical approach \cite{PhysRevB.103.045401,PhysRevB.103.115432} without the correction being free of a divergence and satisfying the Mott relation. However, we have demonstrated that this correction is necessary when using the Kubo formula. This fact is an important guideline when calculating the orbital gravito-ME response in strongly correlated systems using Green's function methods.

In addition, we have classified the intrinsic and extrinsic OGME responses by the magnetic point groups. We have shown that almost all $\mathcal{PT}$-symmetric groups can be detected by the intrinsic part and have no response in the extrinsic part because the orbital magnetic moment vanishes. We have discussed that the extrinsic part can exist even in systems with time-reversal symmetry because dissipation already breaks this symmetry. The symmetry table of the OGME is very similar to the table of the nonlinear Hall effect. 
Thus, the orbital (gravito-) ME effect is also expected in systems where the nonlinear Hall effect occurs.

In experiments, the spin magnetization also contributes to the (gravito-) ME effect. Thus, we need to explore systems with a large orbital magnetization to study the orbital ME effect. Fortunately, phenomena related to orbital magnetic moments, such as the valley Hall effect and the orbital Edelstein effect, have recently been observed in a transition metal dichalcogenide $\mathrm{MoS_2}$ and twisted bilayer graphene with large orbital magnetic moments around the K points. Thus, the extrinsic orbital gravito-ME effect may also be observable in these systems. On the other hand, the intrinsic part becomes dominant in $\mathcal{PT}$-symmetric orbital magnetic orders such as an antiferromagnetic loop current order as discussed in Sec.~\ref{model_calculation}. There, the intrinsic part is strongly enhanced around the Dirac points. This will give an experimental platform for the detection of the intrinsic orbital (gravito-) ME response.

\section*{Acknowledgement}
K.S. acknowledges support as a JSPS research fellow and is supported by JSPS KAKENHI, Grant No.22J23393. R.P. is supported by JSPS KAKENHI No.~JP18K03511.

\bibliography{intrinsicOGME_submit}

\appendix
\begin{widetext}

\section{Detail calculation of the current energy-denstiy correlation function} \label{detail_current-enegy}
The current energy correlation function $\Phi^{i}_{JH}(\q,\omega)$ is given as
\begin{eqnarray}
\Phi^{i}_{JH}(\q,\omega) 
= 
-e \sum_{mn,\kvec} \frac{f(\epsilon_{n\kvec+\q/2}) - f(\epsilon_{m\kvec-\q/2})}{\epsilon_{n\kvec+\q/2} - \epsilon_{m\kvec-\q/2} -( \omega +  i\delta )} 
\bra{u_{m\kvec-\q/2}} v^i_{\kvec} \ket{u_{n\kvec+\q/2}} \frac{\epsilon_{n\kvec+\q/2} + \epsilon_{m\kvec-\q/2}}{2} \braket{u_{n\kvec+\q/2} | u_{m\kvec-\q/2}} . \nonumber \\
\end{eqnarray}
Here, we use the following notations: the eigenenergies are $\epsilon_{n\kvec}$, and the eigenvectors are $\ket{u_{n\kvec}}$ which fulfill $H_{\kvec} \ket{u_{n\kvec}} = \epsilon_{n\kvec} \ket{u_{n\kvec}} $, where $(H_{\kvec} = e^{-i\kvec \cdot \x} (H_0 - \mu N) e^{i \kvec \cdot \x} )$. The velocity operators are defined as $v^i_{\kvec} = \partial H_{\kvec}/ \partial k_i$, and the Fermi distribution function is $f(\epsilon) = 1/(1+ e^{\beta \epsilon})$.
As discussed in the main text, we need to calculate the second-order derivative of the correlation function to obtain the OGME tensor. In the following, we will calculate it for two cases; (A) intraband transitions ($m=n$), and (B) interband transitions ($m \neq n$).

In the case of (A), the correlation function is given as
\begin{eqnarray}
\Phi^{i(\mathrm{A})}_{JH}(\q,\omega) 
= 
-e \sum_{n,\kvec} \frac{f(\epsilon_{n\kvec+\q/2}) - f(\epsilon_{n\kvec-\q/2})}{\epsilon_{n\kvec+\q/2} - \epsilon_{n\kvec-\q/2} -( \omega +  i\delta )} 
\bra{u_{n\kvec-\q/2}} v^i_{\kvec} \ket{u_{n\kvec+\q/2}} \frac{\epsilon_{n\kvec+\q/2} + \epsilon_{n\kvec-\q/2}}{2} \braket{u_{n\kvec+\q/2} | u_{n\kvec-\q/2}} . \nonumber \\
\end{eqnarray}
We expand each coefficient up to the second order of $\q$,
\begin{subequations}
\begin{align}
&\frac{f(\epsilon_{n\kvec+\q/2}) - f(\epsilon_{n\kvec-\q/2})}{\epsilon_{n\kvec+\q/2} - \epsilon_{n\kvec-\q/2} - (\omega + i\delta)}
\simeq 
\frac{-f'_n (\partial_a \epsilon_n) q_a}{\omega + i\delta} - \frac{f'_n (\partial_a \epsilon_{n}) (\partial_b \epsilon_n) q_a q_b}{(\omega + i\delta)^2}
\\
&
\frac{\epsilon_{n\kvec+\q/2} + \epsilon_{n\kvec-\q/2}}{2} 
\simeq
\epsilon_{n\kvec} + 0 \\
&\braket{u_{n\kvec+\q/2} | u_{n\kvec-\q/2}}
\simeq
1 - q_a \braket{u_{n\kvec} | \partial_a u_{n\kvec}} \\
&\bra{u_{n\kvec-\q/2}} v^i_{\kvec}  \ket{u_{n\kvec+\q/2}}
\simeq
\partial_i \epsilon_{n\kvec}
+
\frac{q_a}{2}( \bra{u_{n\kvec}} v^i_{\kvec} \ket{\partial_a u_{n\kvec}}
-
\bra{\partial_a u_{n\kvec}} v^i_{\kvec} \ket{ u_{n\kvec}}
).
\end{align}
\end{subequations}
Here, we define $f'_n = \partial f(\epsilon_{n\kvec})/ \partial \epsilon_{n\kvec}$. Then, the second derivative of $\Phi^{i(\mathrm{A})}_{JH}(\q,\omega)$ for intraband transitions is 
\begin{eqnarray}
\Phi^{i,ab(\mathrm{A})}_{JH}(\omega) q_aq_b
&&=
-e\sum_{n,\kvec} \epsilon_{n\kvec} \biggl[\frac{-f'_n (\partial_a \epsilon_{n\kvec}) }{\omega +i\delta}  \Bigl(  - (\partial_i \epsilon_{n\kvec}) \braket{u_{n\kvec} | \partial_b u_{n\kvec}} 
+
\frac{1}{2} ( \bra{u_{n\kvec}} v^i_{\kvec} \ket{\partial_b u_{n\kvec}}
-
\bra{\partial_b u_{n\kvec}} v^i_{\kvec} \ket{ u_{n\kvec}})
\Bigr) \nonumber \\
&&\hspace{1.8cm}
-\frac{f'_n}{( \omega +i\delta)^2}
(\partial_a \epsilon_{n\kvec}) (\partial_b \epsilon_{n\kvec}) (\partial_i \epsilon_{n\kvec}) \biggr]q_a q_b .
\end{eqnarray}
The term proportional to $1/(\omega + i\delta)^1$ can be transformed as
\begin{eqnarray}
- (\partial_i \epsilon_{n\kvec}) \braket{u_{n\kvec} | \partial_b u_{n\kvec}} 
+
\frac{1}{2} ( \bra{u_{n\kvec}} v^i_{\kvec} \ket{\partial_b u_{n\kvec}}
-
\bra{\partial_b u_{n\kvec}} v^i_{\kvec} \ket{ u_{n\kvec}})
&=&
\frac{1}{2} \Bigl(  
\bra{u_{n\kvec}} v^i_{\kvec} Q_n \ket{\partial_b u_{n\kvec}}
-
\bra{\partial_b u_{n\kvec}} Q_n v^i_{\kvec} \ket{u_{n\kvec}}
\Bigr) \nonumber \\
&=&
\frac{1}{2} \Bigl(
\bra{\partial_i u_{n\kvec}} \epsilon_n -H_{\kvec} \ket{\partial_b u_{n\kvec} } - \mathrm{c.c.}
\Bigr). \nonumber \\
&\equiv& m_n^{ib} - (\mathrm{c.c.}).
\end{eqnarray}
Here, we define $Q_n = 1 - \ket{u_{n\kvec}} \bra{u_{n\kvec}}$.
Using this identity, we can write the correlation function $\Phi^{i,ab(\mathrm{A})}_{JH}(\omega)$ as
\begin{eqnarray}
\Phi^{i,ab(\mathrm{A})}_{JH}(\omega) q_aq_b 
&=&
 -e \sum_{n,\kvec} \epsilon_{n\kvec} \biggl[
\frac{-f'_n}{\omega +i\delta} (\partial_a \epsilon_{n\kvec}) \bigl( m^{ib}_n - m^{bi}_n \bigr)
-
\frac{f'_n}{(\omega + i\delta)^2} (\partial_a \epsilon_{n\kvec}) (\partial_b \epsilon_{n\kvec}) (\partial_i \epsilon_{n\kvec})
\biggr] q_a q_b.
\end{eqnarray}
Finally, we use Eq.~(\ref{beta}) and obtain the contribution from the current energy-density correlation function to the OGME tensor as
\begin{eqnarray}
2i \beta^{JH\mathrm{(A)}}_{li}
&=&
\frac{2}{3} \varepsilon_{ijk} \Phi^{k,lj\mathrm{(A)}}_{JH}(0) \nonumber \\
&=&
-\frac{e}{\delta} \sum_{n,\kvec}  \epsilon_{n\kvec} f'_n (\partial_l \epsilon_{n \kvec}) m^i_n . 
\end{eqnarray}
Here, $\bm{m}_n = \I [\bra{ \bm{\nabla} u_{n\kvec}} \times (\epsilon_{n\kvec} - H_{\kvec})  \ket{\bm{\nabla} u_{n\kvec}}]/2$ is the orbital magnetic moment. This tensor has a Drude-like singularity and originates from the Fermi surface. It is the extrinsic response and is called the Edelstein effect.

Next, we consider the case of interband transitions (B). In this case, the denominator has no singularity. Thus, we can take the limits $\omega, \delta \to 0$. 
The correlation function is given as
\begin{eqnarray}
\Phi^{i(\mathrm{B})}_{JH}(\q,\omega) 
= 
-e \sum_{m\neq n,\kvec} \frac{f(\epsilon_{n\kvec+\q/2}) - f(\epsilon_{m\kvec-\q/2})}{\epsilon_{n\kvec+\q/2} - \epsilon_{m\kvec-\q/2}} 
\bra{u_{m\kvec-\q/2}} v^i_{\kvec} \ket{u_{n\kvec+\q/2}} \frac{\epsilon_{n\kvec+\q/2} + \epsilon_{m\kvec-\q/2}}{2} \braket{u_{n\kvec+\q/2} | u_{m\kvec-\q/2}} . \nonumber \\
\end{eqnarray}
Expanding each coefficients by $\q$ up to the second order,
\begin{subequations}
\begin{align}
&\braket{u_{n\kvec+\q/2} | u_{m\kvec-\q/2}}
\simeq
-q_a \braket{u_{n\kvec} | \partial_a u_{m\kvec}}
-\frac{q_a q_b}{2} \braket{ \partial_a u_{n\kvec} | \partial_b u_{m\kvec}} \\
&\frac{\epsilon_{n\kvec+\q/2} + \epsilon_{m\kvec-\q/2}}{2}
\simeq
\frac{\tilde{\epsilon}_{nm\kvec}}{2} + \frac{q_a}{4} \partial_a \epsilon_{nm\kvec} \label{expand_energy}
\\
&\bra{u_{m\kvec-\q/2}} v^i_{\kvec} \ket{u_{n\kvec+\q/2}}
\simeq
i \epsilon_{mn\kvec} \A^i_{mn}
-\frac{q_a}{2}( \bra{\partial_a u_{m\kvec}} v^i_{\kvec} \ket{u_{n\kvec}}
- \bra{u_{m\kvec}} v^i_{\kvec} \ket{\partial_a u_{n\kvec}}
) \nonumber \\
&\hspace{102pt}=
- \epsilon_{mn\kvec} \braket{u_{m\kvec} | \partial_i u_{n\kvec}}
-\frac{q_a}{2} \Bigl(
- \partial_i \tilde{\epsilon}_{nm\kvec} \braket{u_{m\kvec} | \partial_a u_{n\kvec}} \nonumber \\
&\hspace{114pt}  
-
\sum_{l(\neq n)} \braket{\partial_a u_{m\kvec} | u_{l\kvec}}  \epsilon_{ln\kvec} \braket{u_{l\kvec} | \partial_i u_{n\kvec}}
+ \sum_{l (\neq m)}  \epsilon_{ml\kvec} \braket{u_{m\kvec} | \partial_i u_{l\kvec}} \braket{u_{l\kvec} | \partial_a u_{n\kvec}}
\Bigr) \\
&\frac{f(\epsilon_{n\kvec+\q/2}) - f(\epsilon_{m\kvec-\q/2})}{\epsilon_{n\kvec+\q/2} - \epsilon_{m\kvec-\q/2}}
\simeq
\frac{f_{nm}}{\epsilon_{nm\kvec}}
+
\frac{q_a}{2 \epsilon_{nm\kvec}} \biggl( 
\partial_a \tilde{f}_{nm} - \frac{(\partial_a \tilde{\epsilon}_{nm\kvec}) f_{nm}}{\epsilon_{nm\kvec}}
\biggr).
\end{align}
\end{subequations}
Here, we define $\epsilon_{nm \kvec} = \epsilon_{n\kvec} - \epsilon_{m\kvec}$, $\tilde{\epsilon}_{nm \kvec} = \epsilon_{n\kvec} + \epsilon_{m\kvec}$, $f_{nm \kvec} = f(\epsilon_{n\kvec}) - f(\epsilon_{m\kvec})$ and $\tilde{f}_{nm \kvec} = f(\epsilon_{n\kvec}) + f(\epsilon_{m\kvec})$. $\A^i_{mn} = i \braket{u_{m\kvec} | \partial_i u_{n\kvec}}$ is the Berry connection.
To simplify the calculation, we split the second-order derivative into two cases with respect to Eq.~(\ref{expand_energy}); (i) the contribution from the first term and (ii) the contribution from the second term. In the case of (i), the second derivative of the correlation function is
\begin{eqnarray}
\Phi^{i,ab(\mathrm{B-i})}_{JH} q_a q_b
&=&
-e
\sum_{m \neq n \kvec} \frac{\tilde{\epsilon}_{nm\kvec}}{2} \biggl\{
-
\frac{f_{nm}}{2} \braket{u_{m\kvec} | \partial_i u_{n\kvec}} \braket{\partial_a u_{n\kvec} | \partial_b u_{m\kvec}} 
+
\frac{f_{nm}}{2 \epsilon_{nm\kvec}} \braket{u_{n\kvec} | \partial_a u_{m\kvec}}
\Bigl(
- \partial_i \tilde{\epsilon}_{nm\kvec} \braket{u_{m\kvec} | \partial_b u_{n\kvec}}
\nonumber \\
&&
-
\sum_{l(\neq n)} \braket{\partial_b u_{m\kvec} | u_{l\kvec}} \epsilon_{ln \kvec} \braket{u_{l\kvec} | \partial_i u_{n\kvec}}
+
\sum_{l(\neq m)} \epsilon_{ml\kvec} \braket{u_{m\kvec} | \partial_i u_{l\kvec}} \braket{ u_{l\kvec} | \partial_b u_{n\kvec}}
\Bigr) \nonumber \\
&&
-\frac{1}{2} \braket{u_{n\kvec} | \partial_a u_{m\kvec}} \braket{u_{m\kvec} | \partial_i u_{n\kvec}}
\biggl(
\partial_b \tilde{f}_{nm} - \frac{(\partial_b \tilde{\epsilon}_{nm\kvec}) f_{nm}}{\epsilon_{nm\kvec}}
\biggr)
\biggr\} q_a q_b \nonumber \\
&=&
-e
\sum_{m \neq n \kvec} \frac{\tilde{\epsilon}_{nm\kvec}}{2} \biggl\{
-
f_n \R  [\braket{u_{m\kvec} | \partial_i u_{n\kvec}} \braket{\partial_a u_{n\kvec} | \partial_b u_{m\kvec}} ]
+
\frac{f_n}{ \epsilon_{nm\kvec}} 
\Bigl(
- \partial_i \tilde{\epsilon}_{nm\kvec} \R [ \braket{u_{n\kvec} | \partial_a u_{m\kvec}} \braket{u_{m\kvec} | \partial_b u_{n\kvec}} ]
\nonumber \\
&&
-
\sum_{l(\neq n)}  \epsilon_{ln\kvec} \R [\braket{u_{n\kvec} | \partial_a u_{m\kvec}} \braket{\partial_b u_{m\kvec} | u_{l\kvec}} \braket{u_{l\kvec} | \partial_i u_{n\kvec}} ] 
\nonumber \\
&&
+
\sum_{l(\neq m)}  \epsilon_{ml\kvec} \R [ \braket{u_{n\kvec} | \partial_a u_{m\kvec}} \braket{u_{m\kvec} | \partial_i u_{l\kvec}} \braket{ u_{l\kvec} | \partial_b u_{n\kvec}} ]
\Bigr) \nonumber \\
&&
- \R[ \braket{u_{n\kvec} | \partial_a u_{m\kvec}} \braket{u_{m\kvec} | \partial_i u_{n\kvec}} ]
\biggl(
\partial_b f_{n} - \frac{(\partial_b \tilde{\epsilon}_{nm\kvec}) f_{n}}{\epsilon_{nm\kvec}}
\biggr)
\biggr\} q_a q_b.
\end{eqnarray}
We split this further into two parts regarding the denominator; (a) $(\epsilon_{nm\kvec})^0$ and (b) $(\epsilon_{nm\kvec})^1$ in the denominator. In the case of (a), collecting the terms whose denominator is $(\epsilon_{nm\kvec})^0$, we obtain
\begin{eqnarray}
\Phi^{i,ab(\mathrm{B-i-a})}_{JH}q_a q_b
&=&
-e \sum_{m\neq n,\kvec} \Biggl[ \frac{\tilde{\epsilon}_{nm\kvec}}{2} \biggl\{
-f_n \R[\braket{u_{m\kvec} | \partial_i u_{n\kvec}} \braket{\partial_a u_{n\kvec} | \partial_b u_{m\kvec}} ]
-
(\partial_b f_n) \R [\braket{u_{n\kvec} | \partial_a u_{m\kvec}} \braket{u_{m\kvec} | \partial_i u_{n\kvec}}]
\nonumber \\
&&
+
f_n \sum_{l(\neq n)}\R [\braket{u_{n\kvec} | \partial_a u_{m\kvec}} \braket{ \partial_b u_{m\kvec} | u_{l\kvec}}   \braket{u_{l\kvec} | \partial_i u_{n\kvec}}]
-
f_n \R[\braket{u_{n\kvec} | \partial_a u_{m\kvec}} \braket{u_{m\kvec} | \partial_i u_{n\kvec}}   \braket{u_{n\kvec} | \partial_b u_{n\kvec}}] \biggr\} \nonumber \\
&&
+
\frac{f_n}{2} (\partial_i \tilde{\epsilon}_{nm\kvec}) \R [\braket{u_{n\kvec} | \partial_a u_{m\kvec}} \braket{u_{m\kvec} | \partial_b u_{n\kvec}}]
-
\frac{f_n}{2} (\partial_b \tilde{\epsilon}_{nm\kvec}) \R [\braket{u_{n\kvec} | \partial_a u_{m\kvec}} \braket{u_{m\kvec} | \partial_i u_{n\kvec}}] \nonumber \\
&&
+
\frac{f_n}{2} \sum_{l(\neq n)} \epsilon_{lm\kvec} \R [\braket{u_{n\kvec} | \partial_a u_{m\kvec}} \braket{\partial_b u_{m\kvec} | u_{l\kvec}} \braket{u_{l\kvec} | \partial_i u_{n\kvec}}] \nonumber \\
&&
+
\frac{f_n}{2} \sum_{l(\neq n)} \epsilon_{lm\kvec} \R [\braket{u_{n\kvec} | \partial_a u_{m\kvec}} \braket{u_{m\kvec} |\partial_i  u_{l\kvec}} \braket{u_{l\kvec} | \partial_b u_{n\kvec}}] \Biggr] q_a q_b \nonumber \\
&=&
-e \sum_{m \neq n, \kvec} \frac{f_n}{2} \biggl\{
-\tilde{\epsilon}_{nm\kvec}\R[\braket{u_{m\kvec} | \partial_i u_{n\kvec}} \braket{\partial_a u_{n\kvec} | \partial_b u_{m\kvec}} ]
+ \tilde{\epsilon}_{nm\kvec} \partial_b \R [\braket{u_{n\kvec} | \partial_a u_{m\kvec}} \braket{u_{m\kvec} | \partial_i u_{n\kvec}}] \nonumber \\
&&
+ \sum_{l (\neq n)} \tilde{\epsilon}_{nl\kvec}  \R [\braket{u_{n\kvec} | \partial_a u_{m\kvec}} \braket{ \partial_b u_{m\kvec} | u_{l\kvec}}   \braket{u_{l\kvec} | \partial_i u_{n\kvec}}]
- \tilde{\epsilon}_{nm\kvec} \R[\braket{u_{n\kvec} | \partial_a u_{m\kvec}} \braket{u_{m\kvec} | \partial_i u_{n\kvec}}   \braket{u_{n\kvec} | \partial_b u_{n\kvec}}] \nonumber \\
&&
+ (\partial_i \tilde{\epsilon}_{nm\kvec}) \R [\braket{u_{n\kvec} | \partial_a u_{m\kvec}} \braket{u_{m\kvec} | \partial_b u_{n\kvec}}] 
+
\sum_{l(\neq n)} \epsilon_{lm\kvec} \R [\braket{u_{n\kvec} | \partial_a u_{m\kvec}} \braket{u_{m\kvec} |\partial_i  u_{l\kvec}} \braket{u_{l\kvec} | \partial_b u_{n\kvec}}]
\biggr\} q_a q_b. \nonumber \\
\end{eqnarray}
The third term in this equation can be transformed as
\begin{eqnarray}
&&\sum_{l (\neq n), m(\neq n)} \tilde{\epsilon}_{nl\kvec}  \R [\braket{u_{n\kvec} | \partial_a u_{m\kvec}} \braket{ \partial_b u_{m\kvec} | u_{l\kvec}}   \braket{u_{l\kvec} | \partial_i u_{n\kvec}}] \nonumber \\
&=&
\sum_{l(\neq n)} \tilde{\epsilon}_{nl\kvec} \Bigl\{
\R [\braket{\partial_a u_{n\kvec} | \partial_b u_{l\kvec}} \braket{u_{l\kvec} | \partial_i u_{n\kvec}} ]
- \R [ \braket{\partial_a u_{n\kvec} | u_{n\kvec}} \braket{u_{n\kvec} | \partial_b u_{l\kvec}} \braket{u_{l\kvec} | \partial_i u_{n\kvec}} ]
\Bigr\}.
\end{eqnarray}
Therefore, this term cancels out the first and fourth terms. Finally, we get
\begin{eqnarray}
\Phi^{i,ab(\mathrm{B-i-a})}_{JH}q_a q_b
&=&
-e \sum_{n\neq m,\kvec} \biggl\{
\frac{\tilde{\epsilon}_{nm\kvec} f_n}{2} (-\partial_b \R [\A^a_{nm} \A^i_{mn}])
-
f_n \sum_{l(\neq n)} \R [\A^a_{nm} V^i_{ml,n} \A^b_{ln}]
\biggr\} q_a q_b.
\end{eqnarray}
Here, we define $V^i_{lm,n} = \frac{1}{2} (v^i_{lm} + v^{0i}_n \delta_{lm})$. In the case of (b), collecting the terms whose denominator is $(\epsilon_{nm\kvec})^1$,
\begin{eqnarray}
\Phi^{i,ab(\mathrm{B-i-b})}_{JH} q_a q_b
&=&
-e \sum_{m\neq n,\kvec} \frac{\epsilon_{n\kvec} f_n}{\epsilon_{nm\kvec}} \biggl\{
-
(\partial_i \tilde{\epsilon}_{nm\kvec}) \R [\braket{u_{n\kvec} | \partial_a u_{m\kvec}}] \braket{u_{m\kvec} | \partial_b u_{n\kvec}}
+
(\partial_b \tilde{\epsilon}_{nm\kvec}) \R [\braket{u_{n\kvec} | \partial_a u_{m\kvec}}] \braket{u_{m\kvec} | \partial_i u_{n\kvec}} \nonumber \\
&&
-
\sum_{l(\neq n)} \epsilon_{lm\kvec} \R [\braket{u_{n\kvec} | \partial_a u_{m\kvec}} \braket{\partial_b u_{m\kvec} | u_{l\kvec}} \braket{u_{l\kvec} | \partial_i u_{n\kvec}}] \nonumber \\
&&
+
\sum_{l(\neq n)} \epsilon_{ml\kvec} \R [\braket{u_{n\kvec} | \partial_a u_{m\kvec}} \braket{u_{m\kvec} |\partial_i  u_{l\kvec}} \braket{u_{l\kvec} | \partial_b u_{n\kvec}}]
\biggr\} q_a q_b \nonumber \\
&=&
-e \sum_{m\neq n,\kvec} \frac{2 \epsilon_{n\kvec} f_n}{\epsilon_{nm\kvec}} \sum_{l(\neq n)} \biggl\{
- \R [\A^a_{nm} V^b_{ml,n} \A^i_{ln}] + \R [\A^a_{nm} V^i_{ml,n} \A^b_{ln}]
\biggr\} q_a q_b.
\end{eqnarray}
Next, calculating the case of (ii), we get
\begin{eqnarray}
\Phi^{i,ab(\mathrm{B-ii})}_{JH} q_a q_b
&=&
-e \sum_{m\neq n,\kvec} \frac{f_n}{2} (\partial_b \epsilon_{nm\kvec}) \R [\A^a_{nm} \A^i_{mn}] q_a q_b.
\end{eqnarray}
Collecting all terms, we obtain
\begin{eqnarray}
\Phi^{i,ab(\mathrm{B})}_{JH} q_a q_b
&=&
-e \sum_{n,\kvec} \biggl[
\epsilon_{n\kvec} f_n \sum_{m (\neq n)} \Bigl\{ -\partial_b \R [\A^a_{nm} \A^i_{mn}]
+
\frac{2}{\epsilon_{nm\kvec}} \sum_{l(\neq n)} \bigl(- \R [\A^a_{nm} V^b_{ml,n} \A^i_{ln}] + \R [\A^a_{nm} V^i_{ml,n} \A^b_{ln}] \bigr)
\Bigr\} \nonumber \\
&&
+ f_n 
\Bigl\{
-\sum_{m(\neq n)}\sum_{l(\neq n)} \R [ \A^a_{nm} V^i_{ml,n} \A^b_{ln} ]
+ \frac{1}{4} \frac{\partial^3 \epsilon_{n\kvec} }{\partial k_i \partial k_a \partial k_b}
- \frac{1}{4} \partial_b v^{ia}_{n}
\Bigr\}
\biggr] q_a q_b.
\end{eqnarray}
Here, we define $v^{ia}_n = \bra{u_{n\kvec}} \partial^2 H_{\kvec} / \partial k_i \partial k_a \ket{u_{n\kvec}}$.
Finally, we use Eq.~(\ref{beta}) and obtain the contribution from the current energy-density correlation function to the OGME tensor as
\begin{eqnarray}
2i \beta^{JH\mathrm{(B)}}_{li}
&=&
\frac{2}{3} \varepsilon_{ijk} \Phi^{k,lj\mathrm{(B)}}_{JH} \nonumber \\
&=&
-e \sum_{n\kvec} \biggl[
\epsilon_{n\kvec} f_n \Bigl\{
- \frac{1}{3} \varepsilon_{ijk} \partial_j g^{lk}_n -  \sum_{m(\neq n)}\frac{2}{\epsilon_{nm\kvec}} \R [\A^l_{nm} M^i_{mn}] \Bigr\}
+ f_n \Bigl\{ \frac{2}{3} \sum_{m(\neq n)} \R [\A^l_{nm} M^i_{mn}] 
- \frac{1}{12} \varepsilon_{ijk} \partial_j v^{kl}_n
\Bigr\}
\biggr]. \nonumber \\
\end{eqnarray}
Here, we use the quantum metric $g_{n}^{lk} = \sum_{m(\neq n)}\R [\A^l_{nm} \A^k_{mn}]$ and $\bm{M}_{nm} = \sum_{l(\neq n)}\bm{V}_{ml,n} \times \bm{\A}_{ln}$.
This equation does not depend on the dissipation in the clean limit. Thus, it can be regarded as the intrinsic part.

\section{Derivation of the Mott relation (Eq.~(\ref{Mott}) and  Eq.~(\ref{eMott}))} \label{derivation_Mott}
The intrinsic OGME tensor and the intrinsic orbital magnetoelectric (OME) tensor at the chemical potential $\mu$ and at the temperature $T$ are defined as
\begin{eqnarray}
\chi^{\mathrm{iOGME}}_{ij}(\mu ,T) &=& e \int_{\mathrm{BZ}} \frac{d^3 k}{(2\pi)^3} \sum_n s(\epsilon_{n\kvec}) W^n_{ij} \\
\chi^{\mathrm{iOME}}_{ij}(\mu,T) &=& -e^2 \int_{\mathrm{BZ}} \frac{d^3 k}{(2\pi)^3} \sum_n f(\epsilon_{n\kvec}) W^n_{ij}.
\end{eqnarray}
Here, $f(\epsilon) = 1/(1+e^{\beta \epsilon})$ is the distribution function at the inverse temperature $\beta =1/T$ and $s(\epsilon) = \epsilon f(\epsilon)/T + \log(1 +e^{-\beta \epsilon})$ is the entropy density. $W^n_{ij}$ represents the wave function part in Eq.~(\ref{intrinsicOGME}).
The OGME tensor can be rewritten as
\begin{eqnarray}
\chi^{\mathrm{iOGME}}_{ij}(\mu ,T)
&=&
e \int_{\mathrm{BZ}} \frac{d^3 k}{(2\pi)^3} \sum_n s(\epsilon_{n\kvec}) W^n_{ij} \nonumber \\
&=&
e \int_{\mathrm{BZ}} \frac{d^3 k}{(2\pi)^3} \int d\epsilon \sum_n s(\epsilon) W^n_{ij} \delta(\epsilon - \epsilon_{n\kvec}) \nonumber \\
&=&
e \int_{\mathrm{BZ}} \frac{d^3 k}{(2\pi)^3} \int d\epsilon \sum_n s(\epsilon) W^n_{ij} \frac{d \Theta(\epsilon - \epsilon_{n\kvec})}{d \epsilon} \nonumber \\
&=&
\frac{-1}{e} \int d\epsilon s(\epsilon) \frac{d \chi^{\mathrm{iOME}}_{ij}(\epsilon+\mu,0)}{d\epsilon} \nonumber \\
&=&
\frac{1}{eT} \int d\epsilon (\epsilon - \mu ) \frac{d f(\epsilon -\mu)}{d\epsilon} \chi^{\mathrm{iOME}}_{ij}(\epsilon,0).
\end{eqnarray}
In the final step, we use partial integration and the identity $ds(\epsilon)/d\epsilon = \beta \epsilon ( df(\epsilon)/d\epsilon)$. 
Here, $\delta(x)$ is the delta function and $\Theta(x)$ is the step function.
This formula is called the Mott relation.

The Mott formula is also established for the extrinsic part. The extrinsic OGME tensor and the extrinsic OME tensor at the chemical potential $\mu$ and at the temperature $T$ are given by
\begin{eqnarray}
\chi^{\mathrm{eOGME}}_{ij}(\mu,T)
&=&
\frac{e}{\delta T} \int_{\mathrm{BZ}} \frac{d^3k}{(2 \pi)^3} \sum_n \epsilon_{n\kvec} \frac{\partial f(\epsilon_{n\kvec})}{\partial \epsilon_{n\kvec}} v^{0i}_n m^j_n \\
\chi^{\mathrm{eOME}}_{ij}(\mu,T)
&=&
\frac{e^2}{\delta} \int_{\mathrm{BZ}} \frac{d^3k}{(2 \pi)^3} \sum_n \frac{\partial f(\epsilon_{n\kvec})}{\partial \epsilon_{n\kvec}} v^{0i}_n m^j_n. 
\end{eqnarray}
The extrinsic OGME tensor can be rewritten as
\begin{eqnarray}
\chi^{\mathrm{eOGME}}_{ij}(\mu,T)
&=&
\frac{e}{\delta T} \int_{\mathrm{BZ}} \frac{d^3k}{(2 \pi)^3} \sum_n \epsilon_{n\kvec} \frac{\partial f(\epsilon_{n\kvec})}{\partial \epsilon_{n\kvec}} v^{0i}_n m^j_n \nonumber \\
&=&
\frac{e}{\delta T} \int_{\mathrm{BZ}} \frac{d^3k}{(2 \pi)^3} \int d\epsilon \sum_n \epsilon \frac{\partial f(\epsilon)}{\partial \epsilon} v^{0i}_n m^j_n \frac{ d \Theta(\epsilon - \epsilon_{n\kvec})}{d \epsilon} \nonumber \\
&=&
\frac{1}{eT} \int d\epsilon \epsilon \frac{\partial f(\epsilon)}{\partial \epsilon}  \chi^{\mathrm{eOME}}_{ij}(\epsilon+\mu,0) \nonumber \\
&=&
\frac{1}{eT} \int d\epsilon (\epsilon -\mu) \frac{\partial f(\epsilon-\mu)}{\partial \epsilon}  \chi^{\mathrm{eOME}}_{ij}(\epsilon,0).
\end{eqnarray}
Therefore, the extrinsic part also satisfies the Mott relation.

\end{widetext}
\end{document}